\documentclass[reprint,
superscriptaddress,
amsmath,
amssymb,
prl,
floatfix,
]{revtex4-2}

\usepackage{float}
\usepackage{graphicx}
\usepackage{dcolumn}
\usepackage{bm}
\usepackage[colorlinks,linkcolor=blue,anchorcolor=blue,citecolor=blue,urlcolor=blue]{hyperref}
\usepackage{longtable}

\newcommand{\eqcite}[1]{Eq.~\eqref{#1}}
\newcommand{\figref}[1]{Fig.~\ref{#1}}

\begin{document}

\title{A nano vacuum gauge based on second-order coherence in optical levitation}

\author{Lyu-Hang Liu}
\author{Yu Zheng}
 \email{bigz@ustc.edu.cn}
\author{Yuan Tian}
\author{Long Wang}
\author{Guang-Can Guo}
\author{Fang-Wen Sun}
 \email{fwsun@ustc.edu.cn}
\affiliation{ CAS Key Laboratory of Quantum Information, University of Science and Technology of China, Hefei 230026, China}
\affiliation{ CAS Center for Excellence in Quantum Information and Quantum Physics, University of Science and Technology of China, Hefei 230026, China}

\date{\today}

\begin{abstract}
Accurate measurement of pressure with a wide dynamic range holds significant importance for various applications.
This issue can be realized with a mechanical nano-oscillator, where the pressure-related collisions with surrounding molecules induce its energy dissipation.
However, this energy dissipation of the nano-oscillator may be overshadowed by other processes.
Here, we apply the second-order coherence analysis to accurately characterize those distinct dissipation processes.
Based on an optically levitated nano-oscillator, we successfully obtain precise measurements of the air pressure surrounding the particles from atmosphere to $7\times10^{-6}$ mbar, over $8$ orders of magnitude.
It proves that the mechanical nano-oscillator is an extremely promising candidate for precision pressure sensing applications.
Moreover, the second-order coherence analysis method on a classical system can pave the way to characterize the dynamic properties of an oscillator, which will benefit microscopic thermodynamics, precision measurement, and macroscopic quantum research.
\end{abstract}

\maketitle

\textit{Introduction.-}Mechanical oscillators, known for their high sensitivity to various physical quantities, have garnered considerable attentions\cite{millen2020,gonzalez2021,bachtold2022}.
As an outstanding optomechanical system, optical levitation is a promising platform for exploring mesoscopic-scale quantum superpositions\cite{aspelmeyer2014,millen2015,delic2020,magrini2021,tebbenjohanns2021,ranfagni2022,kamba2022,vijayan2023,pontin2023_1,piotrowski2023} and precision measurement\cite{li2010,millen2014,hebestreit2018_2,kamba2023}.
Due to its unprecedented decoupling from the environment, it plays a significant role in the field of force\cite{ranjit2016,hempston2017,hebestreit2018_1,blakemore2019_2,blakemore2021}, mass\cite{blakemore2019_1,ricci2019,zheng2020,tian2023}, and acceleration\cite{monteiro2017,monteiro2020} sensing.
Since the particle levitated in optical trap experiences random collisions with surrounding gas molecules, this system shows promise as a platform for precise pressure measurement.

In optical levitation, the displacement power spectral density (PSD), which is the Fourier transform of the first-order autocorrelation function of the oscillator's motion trajectory\cite{chatfield1989}, serves as a fundamental analysis tool to identify the decoherence process.
The primary source of decoherence arises from collisions between trapped particles and air molecules.
In the case of a harmonic oscillator, the width of the PSD peak\cite{berg-sorensen2004} provides information about the decoherence induced by this stochastic process.
However, limitations emerge in PSD analysis when dealing with real mechanical systems due to the presence of additional disturbances, such as nonlinearity in the restoring force\cite{gieseler2013,zheng2020,tian2022,zheng2023} and system instability, which can amplify the decoherence rate of the system.

Drawing inspiration from the second-order coherence analysis of light intensity in quantum optics\cite{walls2008}, we utilize an energy PSD analysis of a classical oscillator, which is squared displacement PSD (sdPSD).
It can discern the distinct components of decoherence.

For an anharmonic oscillator, decoherence induced by thermal noise introduces both energy and phase fluctuations, whereas decoherence arising from nonlinearity and instability primarily manifests as phase fluctuations.
The autocorrelation of energy fluctuation can be separately obtained through the second-order coherence analysis.
Consequently, the distinction between different decoherence processes presents a fundamental tool for diverse studies and applications.

Here, we first employed the sdPSD to verify the feasibility of conducting distinct analyses of different decoherence processes, where the system's energy decoherence is calibrated under the influence of anharmonicity in the optical levitation system.
Simultaneously, the energy decoherence related air damping rate of the levitated nano-oscillator is precisely measured from atmosphere to high vacuum.

We also assess its accuracy and robustness in the presence of nonlinearity and instability.
Hence, compared with other air pressure measurement methods\cite{kuhn2017,blakemore2020}, this approach enables the continuous and absolute measurement of air pressure within the range of $10^{3}$ to $7\times10^{-6}$ mbar, offering a high-accuracy and wide-range solution for gas pressure metrology with a mechanical nano-oscillator.

\begin{figure}[t]
    \centering
    \includegraphics[width=\linewidth]{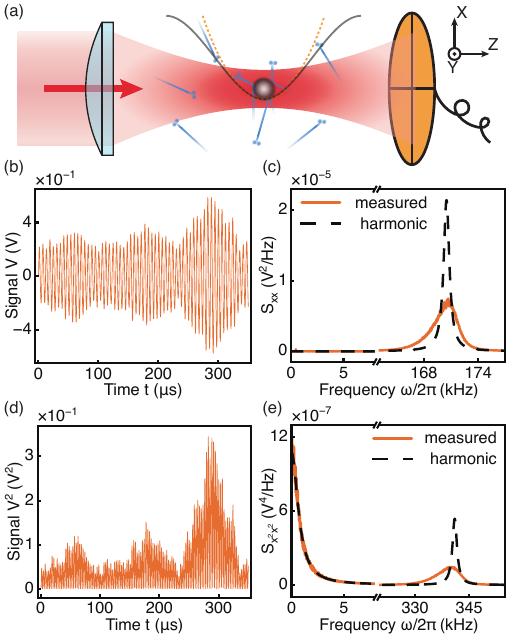}
    \caption{\label{fig:1} (a) The schematic diagram of optical levitation. A single silica nanosphere, which is trapped by a $1064$ nm laser beam, collides with air molecules in thermal motion randomly. (b) Part of the measured photodetector signal along y direction at $1$ mbar. (c) The PSD of the detector signal (orange solid). The dashed black line shows the result from \eqcite{formula1} with experiment condition and harmonic assumption. (d) Part of the squared detector signal along y direction at $1$ mbar. (e) The PSD of the squared detector signal (orange solid). The dashed black line shows the result from \eqcite{formula2}.}
\end{figure}

\textit{Dissipation processes characterized with sdPSD.-}We investigate the measurement of gas pressure using a nano-particle levitated in an optical trap.
The surrounding pressure can be obtained by analyzing the dissipation process of its translational motions.
In a harmonic potential, the PSD of the levitated thermal equilibrium particle's trajectory along one motion degree of freedom can be described as\cite{li2012}
\begin{equation}
    S_{xx}(\omega)=\frac{2k_\mathrm{B}T}{m}\frac{\Gamma_0}{(\omega^2-\omega_0^2)^2+\omega^2\Gamma_0^2}
    \text{,} \label{formula1}
\end{equation}
where $\Gamma_0$ is the damping rate introduced by the collisions with surrounding molecules, which can be used to measure the pressure.
$\omega_0$ is the eigenfrequency, $m$ and $T$ are the mass and temperature of the trapped particle, and $k_\mathrm{B}$ is the Boltzmann constant.
For a harmonic oscillator, $\omega_0$ is a constant.
However, in the presence of the anharmonicity and instability, the eigenfrequency $\omega_0$ is no longer a constant but $\omega_0(x,t)$\cite{gieseler2013}, which induces phase fluctuations.
And \eqcite{formula1} is no longer applicable for the description of the oscillator's frequency domain properties.
Previous studies have attempted to figure out the PSD shape under nonlinearity\cite{suassuna2021}.
However, accurately estimating the PSD of realistic levitation systems is challenging due to the complexity arising from nonlinearity and stability condition\cite{yasuhiro1996,novotny2012}.
\textcolor{black}{In addition to employing PSD, the measurement of $\Gamma_0$ for nonlinear oscillators can be conducted using either the transient trajectories method\cite{flajsmanova2020} or the ring-down method\cite{stipe2001,Di2020}. However, the compatibility of transient trajectories method under high vacuum conditions remains unverified, while the ring-down method is not applicable to optical levitation system with shallow trapping potential well.}

Turning to the perspective of energy, we apply the second-order correlation to present the potential energy PSD analysis.
According to the research on a Langevin oscillator\cite{mishin2016}, we can have the potential energy PSD of a harmonic oscillator without the request of quasi-static assumption.
Since the potential energy is $E_\text{p}=1/2m\omega_0^2x^2$, the PSD of the squared displacement is utilized for convenience, which is
\begin{equation}
    S_{x^2x^2}(\omega)=\frac{4\Gamma_0(k_\mathrm{B}T)^2}{m^2\omega_0^2}\frac{4\Gamma_0^2+\omega^2+4\omega_0^2}{(\omega^2+\Gamma_0^2)[4\omega^2\Gamma_0^2+(\omega^2-4\omega_0^2)^2]}
    \text{.} \label{formula2}
\end{equation}
This sdPSD exhibits two peaks: a zero-frequency peak near $\omega=0$ and a double-frequency peak near $\omega=2\omega_0$. The zero-frequency peak characterizes energy fluctuations due to stochastic energy exchanges with the thermal bath, showing the dissipation process ($\Gamma_0$) from the collisions.

According to the dynamics of the levitated particle's mechanical energy\cite{gieseler2015}, it is worth pointing out that under the quasi-static assumption, where the energy dissipation time is much longer than the oscillation period, the energy fluctuation introduced by the potential's nonlinearity and instability becomes negligible.
And the PSD of the potential energy can be simplified to
$\displaystyle S^{0}_{x^2x^2}(\omega)=\frac{(k_\mathrm{B}T)^2}{m^2\omega_0^4}\frac{\Gamma_0}{\omega^2+\Gamma_0^2}$
{\color{black}{\cite{mishin2016}}},
\textcolor{black}{which has the same shape with the results of mechanical energy PSD in the presence of nonlinearity\cite{SM}}.
It is independent of eigenfrequency $\omega_0(x,t)$ and has been used to study ultra-high Q oscillators\cite{stipe2001,Di2020,pontin2020,bullier2021,golokolenov2023}.

\textit{Experimental measurement of $\Gamma_0$.-}In the optical levitation system, a silica nanosphere (nominal radius $82\pm10$ nm, Bangs labs Inc.) is trapped in a vacuum chamber with an optical potential, which is formed by a focused $1064$ nm Gaussian beam laser\cite{zheng2019_1} with an objective lens (NA=$0.9$), as shown in \figref{fig:1}(a).
Air molecules in the chamber collide randomly with the nanosphere.
The optical potential has Duffing nonlinearity component compared with harmonic one, which results in shift in the oscillator's eigenfrequency at varying amplitudes\cite{gieseler2013}.
The x-axis is parallel to the direction of laser polarization.

The $3$-dimensional motion of the trapped particle is measured with homodyne detection of forward scattering light with $3$ sets of balanced-photodetector\cite{zheng2023}.

We record the trajectories of the nanosphere's thermal motion from atmosphere to high vacuum.
As shown in \figref{fig:1}(c), the measured PSD peak\cite{berg-sorensen2004} is asymmetric and wider than the theoretical fitting from \eqcite{formula1} because of nonlinearity induced significant asymmetric frequency shift.
As for the sdPSD, it manifests as two distinct peaks, as shown in \figref{fig:1}(e).
The double-frequency peak undergoes broadening for the same reason as the PSD.
In contrast, the zero-frequency peak remains unaffected by nonlinearity and aligns consistently with theoretical expectations, which is used to get the value of $\Gamma_0$ with high accuracy in high vacuum.

For comparison, both the PSD and sdPSD of these trajectories are utilized to obtain $\Gamma_0$.
To get the zero-frequency peak of the sdPSD, a low-pass fitting filter is employed by setting a threshold value equal to $1\%$ of the maximum of the sdPSD.
As the value of the sdPSD decreases with increasing frequency from $0$ Hz, we mark a cutoff frequency $\omega_\text{end}$ when the sdPSD value falls below the designated threshold for the first time.
Only the sdPSD data from $0$ Hz to $\omega_\text{end}$ are used for fitting the air damping rate.

\begin{figure}[t]
    \centering
    \includegraphics[width=\linewidth]{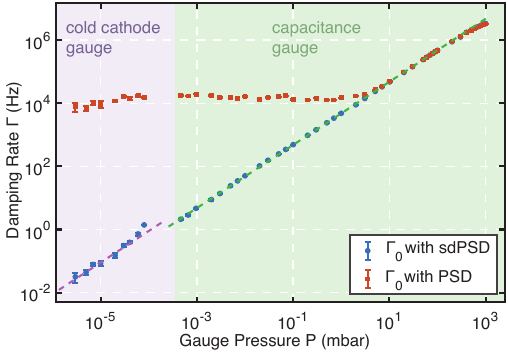}
    \caption{\label{fig:2} The damping rate from \eqcite{formula1} and \eqcite{formula2} as a function of gauge pressure. Each data point and error bar represents the averaged damping rate and standard deviation over $3$ axes and $10$ sampling trajectories. The dashed line is a linear fitting. The duration for each trajectory is $2000$ s.}
\end{figure}

Also, the data of sdPSD at \textcolor{black}{$\omega/2\pi=0\text{Hz}$} is discarded, because the mean value of the squared displacement is non-zero.
.
Figure \ref{fig:2} displays $\Gamma_0$ fitted with \eqcite{formula1} and \eqcite{formula2}, respectively.
PSD fitting remains approximately constant as the air pressure falls below a particular threshold.
However, in this regime, the air damping rate of the particle in vacuum is expected to exhibit a linear relationship with air pressure\cite{epstein1924,li2003}.
This is confirmed with the observation that $\Gamma_0$ derived from sdPSD maintains a linear correlation with air pressure even under high vacuum conditions.

\begin{figure}[t]
    \centering
    \includegraphics[width=\linewidth]{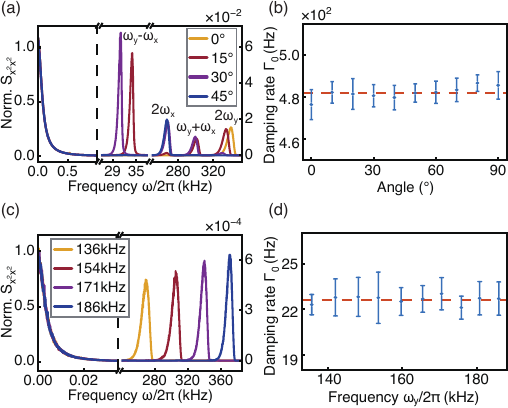}
    \caption{\label{fig:4} (a) sdPSDs of the y-axis trajectories under different polarization angles at $0.1$ mbar. sdPSDs are normalized by the zero-frequency peaks. (b) Damping rates measured along the y-axis under varying polarization angles. (c) sdPSDs of the y-axis trajectories under varying trapping laser powers at $0.005$ mbar.
    sdPSDs are normalized by the zero-frequency peaks. (d) Damping rates measured along the y-axis under varying eigen-frequencies. Error bars in both (b) and (d) represent the standard deviation of ten $300$-second trajectories.}
\end{figure}

Besides the anharmonicity, the instability of the trapping laser constitutes a significant source of error in estimating the parameters related to PSD fitting.
Similar to the nonlinearity that induces PSD broadening, power fluctuations of the trapping laser directly lead to fluctuations in the oscillator's eigenfrequency.
This, in turn, introduces additional broadening to the PSD, resulting in errors in damping rate fitting.
In order to verify the sdPSD's robustness in measuring air damping rates under system instability, we perform the damping rate measurements at various laser intensities and with different polarization directions.
\figref{fig:4}(a) illustrates the observation of sdPSD in a fixed y-axis measurement direction while rotating the polarization direction of the trapping laser using a half-wave plate.
The original double-frequency peak transforms into four peaks, each corresponding to doubled eigenfrequencies along perpendicular and parallel polarization directions, the sum frequency, and the difference frequency.
Meanwhile, the shape of the zero-frequency peak remains unchanged.
The damping rates obtained by fitting the zero-frequency peak at various polarization angles exhibit consistency.
Furthermore, as shown in \figref{fig:4}(c), the frequencies of sdPSD's double-frequency peak shift when changing the intensity of the trapping laser.
However, the damping rate derived from fitting the zero-frequency peaks remains constant.

\textit{A nano vacuum gauge.-}We can apply the high accurate $\Gamma_0$ to measure the surrounding pressure.
In \figref{fig:2} there is a gap obtained through sdPSD.
This gap coincides with the switching from capacitance gauges (CMR361 and CMR364 by Pfeiffer Vacuum GmbH) to the cold cathode gauge (PKR361 by Pfeiffer Vacuum GmbH).
Capacitance gauges are accurate for pressure measurements, whereas the reliability of the cold cathode gauge remains uncertain.
For a cold cathode gauge, apart from the $30\%$ measurement error mentioned in the manufacturer's manual, electrode contamination, such as the carbonization of residual organic molecules, can significantly underestimate the measured pressure compared to the actual values.
In our experimental device, the cold cathode gauge indicated pressure values approximately $50\%$ lower than those recorded by the capacitance gauge under identical conditions\cite{SM}.

However, this gap induced by gauges switching indicates that the air damping rate of the levitated nanoparticle can be used to improve the accuracy of air pressure measurements.
A deterministic relationship exists between the air damping rate of the nanoparticle and the air pressure, which is\cite{beresnev1990,li2012}
\begin{equation}
   \Gamma_0=\frac{6\pi\eta R}{m}\frac{0.619}{0.619+\text{Kn}}(1+c_\text{K})
   \text{,} \label{formula3}
\end{equation}
where $c_\text{K} = 0.31\mathrm{Kn}/\left(0.785+1.152\mathrm{Kn}+\mathrm{Kn}^{2}\right)$, $R$ is the radius of the nanosphere, $\eta$ is the viscosity coefficient of atmosphere air, and $\mathrm{Kn} = \overline{l}/R$ is the Knudsen number.
Additionally, $\overline{l} = k_\text{B}T/\sqrt{2}\pi d^{2}P$ is the mean free path of air molecules, where $P$ is the air pressure and $d$ is the collision diameter of air molecules.

\begin{figure}[ht]
    \centering
    \includegraphics[width=\linewidth]{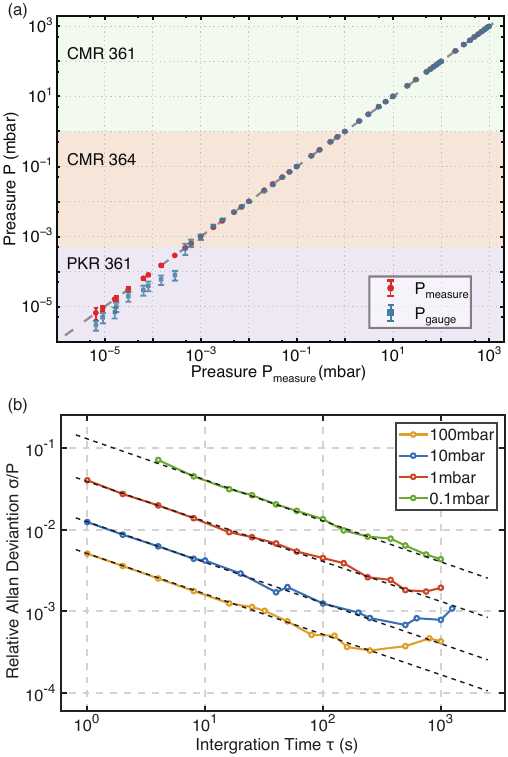}
    \caption{\label{fig:3} (a) Comparison of measured pressures using sdPSD (red dots) and pressure values from gauges (blue squares). The measured pressure data is an average from $3$-axis damping rates, 
    \textcolor{black}{and its error bar represents the standard deviation of $10$ segments of $2000$ s trajectories and the systematic error introduced by the calibration process.}
    The error bar for gauge pressure is sourced from the manufacturer's manuals. (b) Relative Allan deviation analysis of pressure measurements. The black dashed lines represent a curve fitting based on the data in the form of $1/\sqrt{\tau}$.}
\end{figure}

\textcolor{black}{Since it is difficult to accurately determine the radius and mass of a nanoparticle, direct use of \eqcite{formula3} may introduce significant errors. 
For a more accurate pressure measurement, the coefficients in \eqcite{formula3} can be calibrated by using an accurate reference pressure. 
By selecting several damping rates $\Gamma_0$ calculated with sdPSD and corresponding capacitance gauge readings as calibration points, the coefficients $A=6\pi\eta R/m$ and $R$ are obtained by fitting \eqcite{formula3} with the data of calibration points. 
The calibrated \eqcite{formula3} is used in the nanoparticle air pressure measurements. Thus, this calibration process introduces a systematic error equal to the error of the reference capacitance gauge ($0.2\%$ in the manufacturer's manual).
}
Figure \ref{fig:3}(a) demonstrates that employing a single levitated nanosphere enables pressure measurements spanning over eight orders of magnitude, ranging from atmospheric pressure to  $7\times10^{-6}$ mbar,
\textcolor{black}{where the measured pressure with nanoparticle $P_{\text{measure}}$ is extracted with the damping rate $\Gamma_0$ in \figref{fig:2} and \eqcite{formula3}}.
Within the pressure range suitable for capacitance gauge operation, the air pressure values derived from sdPSD-based damping rate measurements align with the gauge readings.
Conversely, within the range of operation for the cold cathode gauge, the pressures obtained from sdPSD measurements higher than the cold cathode gauge.
This discrepancy can be attributed to the measurement error of the cold cathode gauge.

Additionally, the relative Allan deviation for sdPSD pressure measurements is obtained.
Figure \ref{fig:3}(b) illustrates that the relationship among the relative standard deviation ($\sigma$) of pressure measurements, the integration time $\tau$, and the pressure ($P$) follows $\sigma/P=C/\sqrt{P\tau}$, where $C=0.0399(5)\text{ }\mathrm{mbar}^{1/2}\cdot \mathrm{s}^{1/2}$.
We can have the sensitivity of the pressure measurement, which is $S_P=C\sqrt{P}$.
Lower measured pressures necessitate longer measurement times to achieve the same relative error.
This is because the sdPSD pressure measurement relies on fitting the width of the sdPSD zero-frequency peak, which narrows as pressure decreases.
Consequently, longer integration times are necessary for the desired spectral resolution.
In the case of pressure measurements within the operational range of the cold cathode gauge in \figref{fig:3}(a), we utilize $10$ segments of 3-dimensional trajectory data, each segments with a duration of $2000$ s.

The lower boundary of pressure measurements using sdPSD hinges on two constraints.
Firstly, it is determined by the minimum air pressure at which the particle can remain stably trapped under thermal equilibrium without external control.
The application of motion control would introduce an additional damping that alters the width of the sdPSD zero-frequency peak, making the extraction of the air damping rate unfeasible.
Nevertheless, in the absence of feedback control, optically levitated particles are susceptible to loss under conditions of high vacuum\cite{zheng2019_2}.
In our experiments, particle loss occurs during measurements at an air pressure of approximately $5\times10^{-6}$ mbar.
Secondly, when the optically levitated nano-particle is in a vacuum below $10^{-8}$ mbar, recoil heating from the trapping laser has become the dominant energy decoherence source over stochastic collisions with air molecules\cite{gonzalez2021,Jain2016,Kamba2021}, making the sdPSD-based air pressure measurements impractical.

\textit{Conclusion.-}In conclusion, we have presented a novel method for quantifying energy dissipation in levitated mechanical systems.
This method relies on second-order coherence analysis with the squared displacement power spectrum density.
Through this approach, we are able to mitigate the influence of phase decoherence stemming from nonlinearities and instabilities on the calculation of the nano-oscillator's damping rate.
This proposed method for determining the damping rate demonstrates robustness across independent oscillation axes and at different eigenfrequencies.
In addition, we have successfully applied this method to measure the air pressure from the atmosphere to $7\times10^{-6}$ mbar with high accuracy.

The method presented in this study for addressing anharmonic oscillators is versatile and can be applied to a broader spectrum of nano-, micro-mechanical systems.
Additionally, our work offers a reliable means of absolute pressure measurement, particularly under wide dynamic range conditions.
The air pressure value derived using this method directly aligns with the definition of air pressure, which signifies the mechanical impact resulting from air molecule collisions.
This characteristic makes it a valuable reference for calibrating pressure gauges.

This work was supported by the National Natural Science Foundation of China (Grant No. 12104438 and No. 62225506), the CAS Project for Young Scientists in Basic Research (Grant No. YSBR-049), and the Fundamental Research Funds for the Central Universities.



\providecommand{\noopsort}[1]{}\providecommand{\singleletter}[1]{#1}%

\end{document}